\documentclass{article}
\usepackage{spconf,amsmath,graphicx}
\usepackage{multirow}

\title{Comprehensive evaluation of statistical speech waveform synthesis}

\name{\parbox{\linewidth}{\centering
Thomas Merritt, Bartosz Putrycz, Adam Nadolski, Tianjun Ye, Daniel Korzekwa, \\ 
Wiktor Dolecki, Thomas Drugman, Viacheslav Klimkov, Alexis Moinet, Andrew Breen, \\ 
Rafal Kuklinski, Nikko Strom, Roberto Barra-Chicote \thanks{Corresponding authors emails: \{thommer, bartosz, anadolsk, rchicote\}@amazon.com. Paper accepted for SLT 2018.}
}}
\address{Amazon.com}

\begin{document}

\maketitle

\begin{abstract}
Statistical TTS systems that directly predict the speech waveform have recently reported improvements in synthesis quality. This investigation evaluates Amazon's statistical speech waveform synthesis (SSWS) system. An in-depth evaluation of SSWS is conducted across a number of domains to better understand the consistency in quality. The results of this evaluation are validated by repeating the procedure on a separate group of testers. Finally, an analysis of the nature of speech errors of SSWS compared to hybrid unit selection synthesis is conducted to identify the strengths and weaknesses of SSWS. Having a deeper insight into SSWS allows us to better define the focus of future work to improve this new technology.
\end{abstract}


\section{Introduction}
For a long time speech synthesis systems have involved a trade-off between high naturalness and flexibility. Unit selection synthesis \cite{hunt1996unit, taylor1998architecture, taylor2006target} offers extremely high naturalness under best-case conditions, however it is impossible to fully cover the range of different speech sounds in the unit database. Once the optimal units are not available for synthesis the naturalness of unit selection drops dramatically. Conventional statistical parametric speech synthesis (SPSS) with vocoder-derived speech parameters \cite{zen2009statistical, king2011introduction, zen2013statistical, zen2015acoustic} provides speech which is of an extremely stable level of naturalness, however is far short of that of natural speech. Hybrid synthesis \cite{ling2008ustc, yan2010rich, qian2013unified, merritt2016deep} was proposed to bridge the gap between SPSS and unit selection by improving the stability of unit selection. Hybrid synthesis is, however, still grounded in the unit selection paradigm, so whilst it mitigates some of the most extreme effects of unit selection instabilities, it is still limited by the unit database available. 

In conventional vocoder-based SPSS, statistical models are used to predict the distribution of speech parameters. These parameters are then passed through a vocoder \cite{kawahara2006straight, morise2016world, drugman2012deterministic} to produce the speech waveform. The vocoder imposes many manually-crafted assumptions about speech production (for example the source-filter model \cite{fant1970acoustic}). These assumptions result in a drop in naturalness purely from vocoding, before any modelling has taken place \cite{merritt2014investigating, merritt2015attributing}. Recently there has been a shift in the statistical speech synthesis paradigm, resulting in a number of synthesis systems using statistical models to directly predict the speech waveform \cite{tokuda2015directly, oord2016wavenet, arik2017deep, mehri2017an}. As these systems are grounded in the statistical speech synthesis paradigm it is hoped that they retain more flexibility than unit selection-based synthesis methods, thus reducing the compromise required between high quality and flexibility. 

WaveNet is a statistical speech waveform synthesis (SSWS) system, first announced in \cite{oord2016wavenet}. Since this lab report there has been much work in literature on SSWS, demonstrating that this more direct approach to TTS results in improvements in speech naturalness. This work includes the Deep Voice system \cite{arik2017deep,arik2017deep2,ping2017deep3}, the SampleRNN and Char2Wav systems \cite{mehri2017an,sotelo2017char2wav} and further work on WaveNet \cite{oord2017parallel}. The synthesis system investigated in this paper follows on from the recent work in literature. 

In this investigation an SSWS system is built on Amazon speech recordings. Previous published work on SSWS have not produced an in-depth description of the evaluation methods undertaken and also have not provided a further detailed analysis of how the system performs across differing speech domains, compared to conventional synthesis technologies. In this investigation the SSWS system is robustly tested and we demonstrate that our results are repeatable and reliable. In addition a detailed analysis of the shortcomings of the implemented SSWS system is conducted.

\section{Amazon's SSWS system}

\begin{figure*}[ht!]
\centering
\includegraphics[width=120mm]{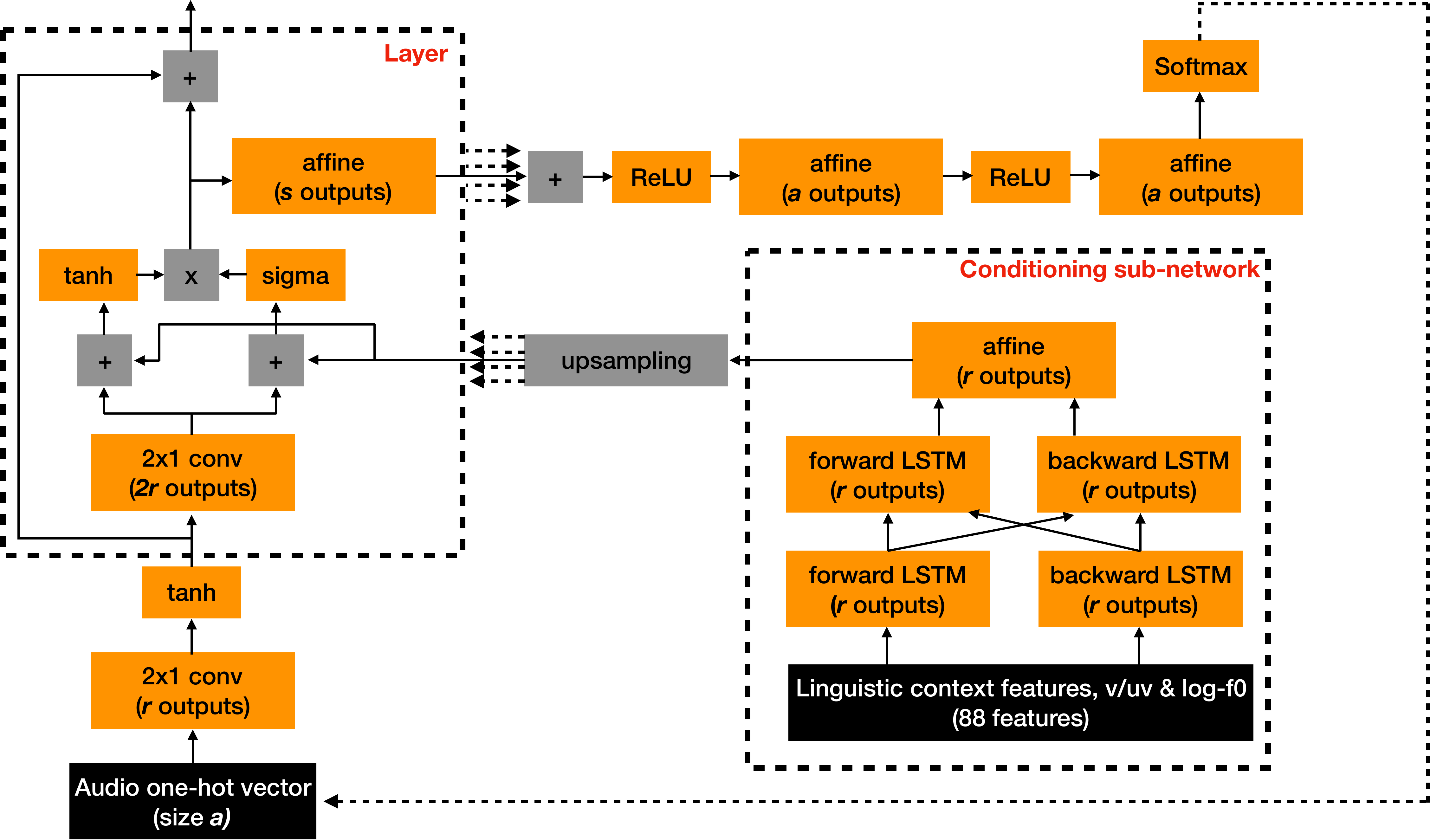}
\caption{Overview of SSWS topology, based on \cite{arik2017deep}. `a' denotes the number of bins used to quantise the audio, `r' denotes the number of residual channels, `s' denotes the number of skip channels. \label{fig:amawave_topology}}
\end{figure*}

The model topology used for the SSWS system is shown in Figure \ref{fig:amawave_topology}. This system is very similar to that described in \cite{arik2017deep}. The differences between the two systems are: 1) here we use LSTMs in the conditioning sub-network instead of quasi-RNNs (QRNNs) \cite{bradbury2016quasi}, 2) instead of interleaving the recurrent output in the conditioning sub-network here we put these activations through an affine transform. 

The number of residual channels used is 128 and the number of skip channels is 1024. The SSWS system used in this investigation is composed of 4 blocks with 10 layers per block. The dilation at each layer is $2^{n-1}$, with $n$ equal to the layer in the current block. The dilations reset at each block (the same approach as in \cite{oord2016wavenet}). By stacking up these dilated convolutions the model is able to construct a wide receptive field of the surrounding speech waveform samples. 

The conditioning sub-network of the SSWS system takes as input the same 86 linguistic context features that provide the input for the SPSS system which forms one of the baseline systems in this investigation. In addition the voiced/unvoiced decision and log-f0 values are input to the conditioning sub-network, these are predicted by the baseline SPSS system. Given these input streams, the conditioning sub-network uses two layers of bi-directional LSTMs, each LSTM has an output size of 128. The output of these LSTMs are stacked and the dimensionality is reduced to 128 (the number of residual channels) using an affine transform. Upsampling is then performed to transform from the frame-level (the unit of time that the conditioning sub-network is operating at) to the sample level (the unit of time that the rest of the SSWS model is operating at). This upsampling is done online to enable for joint training of the speech production model and the conditioning sub-network. This means that all of the weights of the network are updated during training, resulting in the weights of the components in the conditioning sub-network being updated to ensure that the conditioning embedding is as effective as possible for the task of providing context to the rest of the SSWS network.

For this investigation 24kHz audio was used, quantised into 1024 bins (10-bits) using $\mu$ law encoding. The audio was split into chunks of 165 vocoder frames. Of these 165 frames, 120 frames is the content to be used for training. There are 35 frames before the content that are included to add historic context for the SSWS network. These frames `warm up' the network so that the receptive field of the network has some context before we begin updating the weights. Likewise, there are 10 frames of future context added after the content frames. This is added to give the bi-directional LSTMs in the conditioning sub-network context of the future for when we begin updating the weights. The weights of the SSWS network are not updated during the historic or future frames. These are purely included to give the network the context that it needs to converge.

An initial learning rate of $5*10^{-4}$ was used with an annealing rate of 0.836 applied to the learning rate after each epoch. Adam optimiser \cite{kingma2014adam} was used ($\beta_{1}=0.9$, $\beta_{2}=0.999$ and $\epsilon=10^{-8}$).

At synthesis-time a Gumbel noise component is added to introduce an element of noise when sampling from the softmax of the SSWS system \cite{jang2016categorical}. Synthesis was conducted in the same chunks of 165 vocoder frames as was used to train the network, with the samples predicted by the network in the previous chunk used at the input of the historic context `warm up' frames.

\section{Evaluation}

\subsection{Baseline systems}
There were two baseline systems included in the perceptual testing in this investigation (Table \ref{tab:systems}). The first of these systems is `hybrid'. This is a unit selection system that is driven by predictions from a state-level statistical parametric model (as used in \cite{klimkov2017phrase}). 

The second of the baseline systems is `SPSS'. This is a conventional DNN-based statistical parametric speech synthesis system that predicts log-f0, voiced/unvoiced decision, band-aperiodicities and mel-cepstra speech parameters. These parameters are extracted using the WORLD vocoder \cite{morise2016world}. At synthesis-time, the parameters are passed through the WORLD vocoder to produce the speech waveform. The log-f0 and voiced/unvoiced features output from this SPSS system are used to condition the SSWS synthesis system.

\begin{table} [htbp!]
\footnotesize
\begin{center}
\caption{Systems present in listening tests.}
\label{tab:systems}
\begin{tabular}{|c|c|}
\hline
ID & Description \\ \hline \hline
SSWS & The system proposed in this investigation \\ \hline
hybrid & Unit selection system driven by state-level \\
 & statistical parametric predictions \\ \hline 
recordings & Natural speech waveforms \\ \hline
SPSS & Conventional statistical parametric speech synthesis \\
 & system with WORLD vocoder \cite{morise2016world} \\ \hline
\end{tabular}
\end{center}
\end{table}

\subsection{MUSHRA evaluation}
All systems present in the perceptual evaluation were trained on around 20 hours of speech from a female US-English voice. The listening test was conducted based on the MUSHRA (MUltiple Stimuli with Hidden Reference and Anchor) methodology \cite{itura2003mushra}. This method of perceptual testing, originally developed to evaluate audio codecs, has been found to be very powerful at detecting differences between speech synthesis systems \cite{henter2014measuring, watts2016hmms}. In MUSHRA, subjects are asked to rate systems on a scale from 0 to 100. All systems in the test are presented side-by-side to the listener on the same screen, for the same utterance.

Usually in the MUSHRA paradigm upper and lower anchors are present in the test and listeners are instructed to find these and rate them as 100 and 0 respectively. In the field of speech synthesis, however, it is extremely difficult to define a robust lower anchor (i.e., a system that is guaranteed to be always worse than the other systems in the test), therefore in the MUSHRA tests conducted in this investigation no lower anchor was used (as was also the case in \cite{henter2014measuring, merritt2016deep, watts2016hmms}). Additionally, when conducting the perceptual tests, listeners were not forced to rate any of the systems as 100.

The MUSHRA test was conducted using 50 listeners all of whom were native US-English speakers with no self-reported hearing impairments. Each listener rates 40 test screens of utterances (every screen contains all 4 systems). A total of 200 test utterances were used in the test. These were split across the 50 listeners to ensure every test utterance was rated 10 times. This number of listeners has been demonstrated to be sufficient to ensure reliable outcomes of perceptual testing \cite{wester2015are}. In order to gauge the naturalness across the large number of domains required for Amazon's speech synthesis the 200 utterances selected for testing came from 9 different speech domains (entertainment, infotainment, texting, accessibility calling, flash-briefing, news, spelling and navigation). The number of utterances per domain is shown in Table \ref{tab:mushra_by_domain}. The domains were split proportionally to each of the listeners to ensure that the speech domains that listeners heard were balanced.

To ensure that the results obtained from the perceptual evaluation are robust and repeatable, the MUSHRA test was run twice using different listeners (100 listeners were recruited in total). This enables for comparison between the two groups of listeners (we refer to them here as ``group A'' and ``group B''). The listeners used for this evaluation were not speech experts. They were recruited to uniformly represent the states in the USA and were also balanced evenly by gender. The listeners conducted the MUSHRA test from their homes using headphones.

\subsection{Analysis of speech errors}
In order to get a better understanding of the changes in quality between the different technologies tested in this investigation, further testing was conducted with Amazon employees that are native speakers of US-English and without self-reported hearing impairments. They participated in office conditions using headphones. The testers were presented with the audio samples produced from the different systems present in the MUSHRA test (all 200 utterances produced by the 4 systems in the MUSHRA test were presented to the linguists) and were asked to flag where issues occur in the synthesised speech and what the nature of the errors were. The categories they could select from were: `audio glitch', `stress', `intonation/prosody', `pronunciation', `incorrect pause insertion', `incorrect pitch insertion', `text normalisation' and `other'. The testers were also asked to rate how severe this speech error was (critical, medium or minor). The objective of this test is to gain some further insight into the nature of the speech issues that occur in the different speech synthesis paradigms.

\section{Results}

\subsection{MUSHRA evaluation}

\subsubsection{Overall evaluation on multiple domains}
\begin{figure}[ht!]
\centering
\includegraphics[width=85mm]{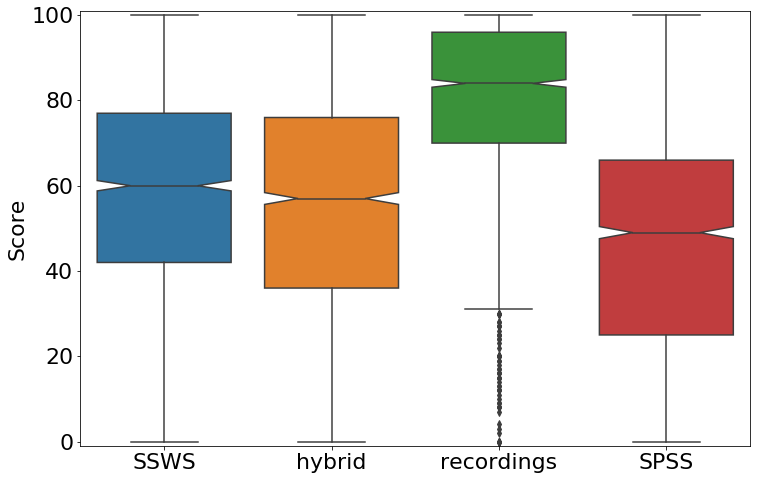}
\caption{Boxplot of absolute values from MUSHRA test on listener group A \label{fig:mushra_abs_group_a}}
\end{figure}

\begin{figure}[ht!]
\centering
\includegraphics[width=85mm]{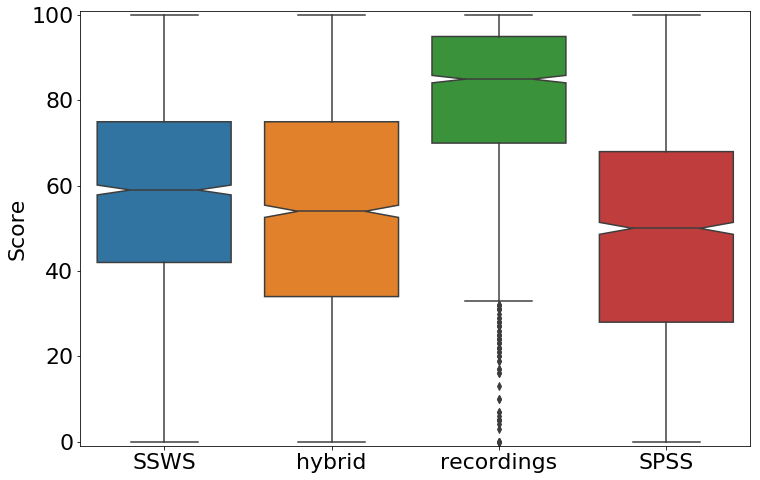}
\caption{Boxplot of absolute values from MUSHRA test on listener group B \label{fig:mushra_abs_group_b}}
\end{figure}

The listener responses from the MUSHRA evaluation are summarised in Table \ref{tab:mushra}. The responses from listener group A are shown in Figure \ref{fig:mushra_abs_group_a}. Responses from listener group B are shown in Figure \ref{fig:mushra_abs_group_b}. For both group A and group B all of the systems present in the MUSHRA test are statistically significant from each other at a p-value of 0.01 in terms of the absolute scores awarded by listeners. For detecting statistical significance a two-sided \mbox{t-test} is used. Holm-Bonferroni correction was applied due to the large number of condition pairs to compare.

\begin{figure}[ht!]
\centering
\includegraphics[width=85mm]{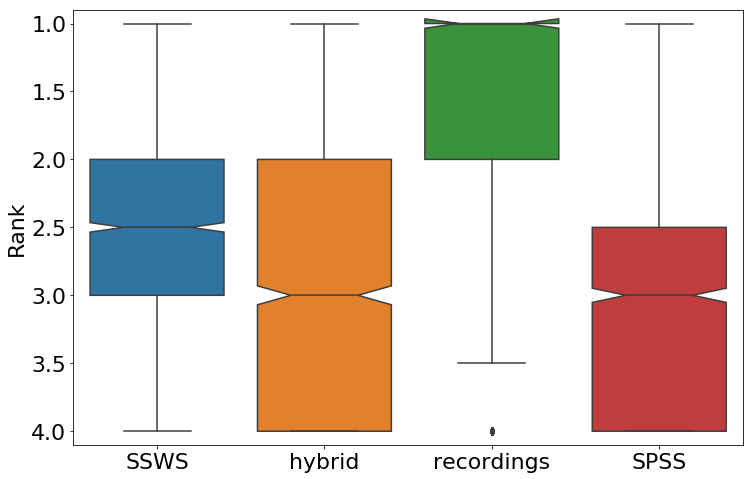}
\caption{Boxplot of the rank order from MUSHRA test on listener group A \label{fig:mushra_rank_group_a}}
\vspace{-4mm}
\end{figure}

\begin{figure}[ht!]
\centering
\includegraphics[width=85mm]{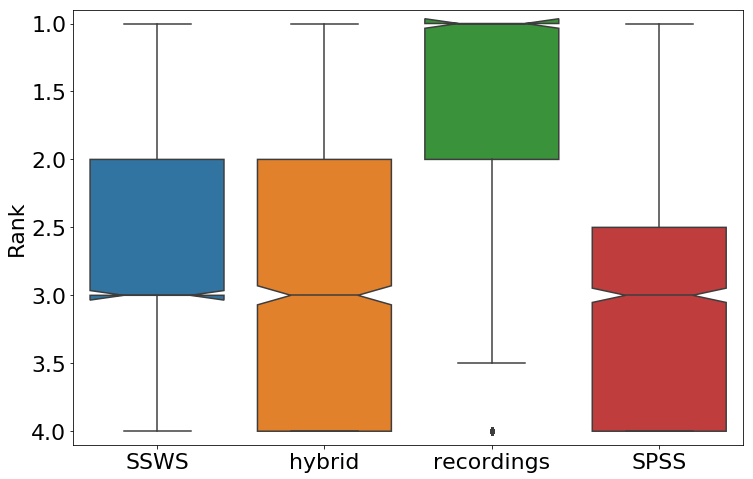}
\caption{Boxplot of the rank order from MUSHRA test on listener group B \label{fig:mushra_rank_group_b}}
\end{figure}

The responses of listener group A processed in terms of the rank order awarded to the synthesis systems are shown in Figure \ref{fig:mushra_rank_group_a}. Responses of listener group B processed in terms of the rank-order awarded are shown in Figure \ref{fig:mushra_rank_group_b}. For both listener groups, A and B, all of the systems present in the MUSHRA test are statistically significant from each other at a p-value of 0.01 in terms of the rank order awarded by listeners. For detecting statistical significance the Wilcoxon signed-rank test is used. Holm-Bonferroni correction was applied.

\begin{table} [htbp!]
\footnotesize
\begin{center}
\caption{Listener ratings from MUSHRA test. Bold typeface denotes the preferred system.}
\label{tab:mushra}
\begin{tabular}{|c|c|c|c|c|c|}
\hline
Listener & System & Mean & Median & Mean & Median \\ 
group & & score & score & rank & rank \\ \hline \hline
 & recordings & 79.56 & 84 & 1.54 & 1 \\ \cline{2-6}
A & SSWS & \textbf{58.24} & \textbf{60} & \textbf{2.57} & \textbf{2.5} \\ \cline{2-6}
 & hybrid & 55.07 & 57 & 2.76 & 3 \\ \cline{2-6}
 & SPSS & 45.46 & 49 & 3.12 & 3 \\ \hline
 & recordings & 79.82 & 85 & 1.49 & 1 \\ \cline{2-6}
B & SSWS & \textbf{57.65} & \textbf{59} & \textbf{2.59} & \textbf{3} \\ \cline{2-6}
 & hybrid & 53.99 & 54 & 2.85 & \textbf{3} \\ \cline{2-6}
 & SPSS & 47.60 & 50 & 3.08 & \textbf{3} \\ \hline
 & recordings & 79.69 & 84 & 1.52 & 1 \\ \cline{2-6}
Combined & SSWS & \textbf{57.95} & \textbf{60} & \textbf{2.58} & \textbf{2.5} \\ \cline{2-6}
(A \& B) & hybrid & 54.53 & 55 & 2.80 & 3 \\ \cline{2-6}
 & SPSS & 46.52 & 50 & 3.10 & 3 \\ \hline
\end{tabular}
\vspace{-4mm}
\end{center}
\end{table}

By observing the results obtained from listener group A and listener group B, two independent sets of listeners, we can see that the outcomes of the MUSHRA test are repeatable. This allows us to conclude that over all of the utterances present in the perceptual test the SSWS system is significantly preferred to the hybrid synthesis system. Additionally both the SSWS and hybrid synthesis systems are significantly preferred to the SPSS system. The test set used in this investigation is made up of multiple speech domains. We will now look more closely into the listener responses to further analyse the performance of the different synthesis systems on each of the speech domains present in the perceptual test.

\subsubsection{Results by speech domain}
Given that the above analysis of listener responses found that the behaviour of the two groups of listeners were very similar (and due to the constraints of space in this report), we will now combine the responses from group A and group B for further analysis of the effect of speech domain on perceived naturalness. The listener responses in terms of absolute values are presented in Table \ref{tab:mushra_by_domain}. From this breakdown by speech domain we can see that SSWS introduces a significant improvement over hybrid synthesis, at a p-value of 0.01, in the domains of: infotainment, calling, news and navigation. There was no statistical significance found between SSWS and hybrid synthesis for the domains of: entertainment, texting, accessibility and flash-briefing. Hybrid synthesis was found to be significantly better than SSWS for the spelling domain. Tests for statistical significance were conducted using a two-sided t-test, applying Holm-Bonferroni correction. 

It is interesting that the only domain where SSWS is found to be significantly worse than hybrid synthesis is in the spelling domain. Informal listening indicates that the main reason behind the preference for hybrid synthesis is due to the unnatural prosody contour in the SSWS samples in this domain. This highlights that the SPSS system which generates the prosody contour used to condition SSWS is particularly ill-suited to the spelling domain, which causes SSWS to subsequently not perform as well as hybrid synthesis. 

\begin{table} [htbp!]
\footnotesize
\begin{center}
\caption{Listener ratings from MUSHRA test by speech domain. Bold typeface denotes the preferred system.}
\label{tab:mushra_by_domain}
\begin{tabular}{|c|c|c|c|c|}
\hline
Domain & System & Mean & Median & p-value: \\ 
(number of & & score & score & SSWS \\
utterances) & & & & vs hybrid \\ \hline \hline
 & recordings & 79.64 & 84 & \multirow{4}{*}{0.3152} \\ \cline{2-4}
Entertainment & SSWS & \textbf{59.45} & \textbf{60} & \\ \cline{2-4}
(25) & hybrid & 57.84 & \textbf{60} & \\ \cline{2-4}
 & SPSS & 46.85 & 50 & \\ \hline
 & recordings & 80.23 & 84 & \multirow{4}{*}{$<$0.0001} \\ \cline{2-4}
Infotainment & SSWS & \textbf{59.02} & \textbf{60} & \\ \cline{2-4}
(25) & hybrid & 51.09 & 50 & \\ \cline{2-4}
 & SPSS & 47.77 & 50 & \\ \hline
 & recordings & 76.97 & 82 & \multirow{4}{*}{0.1752} \\ \cline{2-4}
Texting & SSWS & \textbf{56.37} & \textbf{56} & \\ \cline{2-4}
(25) & hybrid & 54.27 & \textbf{56} & \\ \cline{2-4}
 & SPSS & 45.94 & 50 & \\ \hline
 & recordings & 82.83 & 88 & \multirow{4}{*}{0.8225} \\ \cline{2-4}
Accessibility & SSWS & 57.85 & \textbf{60} & \\ \cline{2-4}
(15) & hybrid & \textbf{58.32} & 59 & \\ \cline{2-4}
 & SPSS & 47.12 & 50 & \\ \hline
 & recordings & 79.59 & 84 & \multirow{4}{*}{0.0005} \\ \cline{2-4}
Calling & SSWS & \textbf{60.25} & \textbf{61} & \\ \cline{2-4}
(25) & hybrid & 54.44 & 55 & \\ \cline{2-4}
 & SPSS & 49.44 & 50 & \\ \hline
 & recordings & 77.84 & 84 & \multirow{4}{*}{0.1606} \\ \cline{2-4}
Flash-briefing & SSWS & \textbf{54.50} & \textbf{55} & \\ \cline{2-4}
(15) & hybrid & 51.63 & 50 & \\ \cline{2-4}
 & SPSS & 42.95 & 44 & \\ \hline
 & recordings & 79.97 & 84 & \multirow{4}{*}{0.0003} \\ \cline{2-4}
News & SSWS & \textbf{60.51} & \textbf{62} & \\ \cline{2-4}
(35) & hybrid & 55.64 & 58 & \\ \cline{2-4}
 & SPSS & 47.58 & 50 & \\ \hline
 & recordings & 81.32 & 89.5 & \multirow{4}{*}{$<$0.0001} \\ \cline{2-4}
Spelling & SSWS & 45.29 & 48.5 & \\ \cline{2-4}
(10) & hybrid & \textbf{57.77} & \textbf{60} & \\ \cline{2-4}
 & SPSS & 37.08 & 34 & \\ \hline
 & recordings & 80.19 & 85 & \multirow{4}{*}{0.0003} \\ \cline{2-4}
Navigation & SSWS & \textbf{58.28} & \textbf{60} & \\ \cline{2-4}
(25) & hybrid & 51.67 & 50 & \\ \cline{2-4}
 & SPSS & 46.66 & 50 & \\ \hline
\end{tabular}
\end{center}
\end{table}

\subsection{Analysis of speech errors}
The errors detected by the testers are shown in Table \ref{tab:ling_analysis}\footnote{Note that these are the perceived `cause' of errors, the labels are self-diagnosed by the listener. The same front-end was used for all synthesis systems tested in this investigation.}. The numbers shown reflect the total number of speech errors detected in the test corpus of 200 utterances. Table \ref{tab:ling_analysis} shows the listeners' responses added together. Due to constraints on space only the errors detected for the SSWS and hybrid synthesis systems are presented here. Table \ref{tab:ling_analysis} shows that the synthesis improvements of the SSWS system over hybrid synthesis come largely from a reduction in the cases of perceived: incorrect pause insertion, incorrect pitch accent, issues with intonation/prosody and issues with text normalisation. The downside of SSWS compared to hybrid synthesis is an increase in the number of perceived audio glitches. Table \ref{tab:swss_audio_glitch_domain} further breaks down these reported audio glitches by domain. Table \ref{tab:swss_audio_glitch_domain} indicates that audio glitches are not associated to any particular domain.

\begin{table} [htbp!]
\footnotesize
\begin{center}
\caption{Perceived cause of errors.}
\label{tab:ling_analysis}
\begin{tabular}{|c|c|c|c|c|}
\hline
System & Category & Critical & Medium & Minor \\ \hline
\multirow{8}{*}{Hybrid} & Audio glitch & 3 & 9 & 12 \\
 & Incorrect pause insertion & 1 & 12 & 9 \\
 & Incorrect pitch accent & 0 & 17 & 18 \\
 & Intonation/Prosody & 1 & 5 & 12 \\
 & Pronunciation & 3 & 13 & 3 \\
 & Stress & 1 & 1 & 2 \\
 & Text normalisation & 1 & 10 & 1 \\
 & Other & 2 & 2 & 6 \\ \hline
\multirow{8}{*}{SSWS} & Audio glitch & 13 & 7 & 20 \\
 & Incorrect pause insertion & 0 & 8 & 4 \\
 & Incorrect pitch accent & 0 & 4 & 13 \\
 & Intonation/Prosody & 0 & 2 & 2 \\
 & Pronunciation & 1 & 12 & 5 \\
 & Stress & 1 & 0 & 0 \\
 & Text normalisation & 1 & 3 & 0 \\
 & Other & 4 & 2 & 8 \\ \hline
\end{tabular}
\end{center}
\end{table}

It is encouraging that many of the types of artefacts that are common problems in hybrid synthesis are mitigated by SSWS. However, future work clearly needs to be focused on increasing the stability of SSWS to reduce the number of audio glitches. 

Through informal listening, we observed that the quality and accuracy of the conditioning of SSWS has a direct impact on the final stability of the model. There have been many investigations published that have implemented differing methods of conditioning the network. These include conditioning the network: as a neural vocoder \cite{hu2017ustc}, using Tacotron \cite{Wang2017, shen2017natural} or using speaker embeddings \cite{arik2017deep2}. These investigations have reported improvements in overall naturalness. However, they have not always reported perceptual investigations into how these conditioning methods affect the stability of the speech waveform model. \cite{shen2017natural} does, however, provide some insight into the nature of the errors experienced with the Tacotron 2 system. Additionally, recent work has looked into the use of mixture of logistics for improving the quantisation of predicted audio for SSWS \cite{oord2017parallel}. However there are no reported perceptual metrics as to how this change affects the stability of the model. Future work should investigate how such changes affect the stability of the model.

\begin{table} [htbp!]
\footnotesize
\begin{center}
\caption{Reported audio glitches in the SSWS system per domain.}
\label{tab:swss_audio_glitch_domain}
\begin{tabular}{|c|c|c|c|}
\hline
Domain & Critical & Medium & Minor \\ \hline
Entertainment & 0 & 1 & 4 \\
Infotainment & 1 & 3 & 4 \\
Texting & 1 & 0 & 1 \\
Accessibility & 2 & 0 & 0 \\
Calling & 0 & 0 & 0 \\
Flash-briefing & 3 & 1 & 6 \\
News & 0 & 0 & 3 \\
Spelling & 3 & 0 & 0 \\
Navigation & 3 & 2 & 2 \\ \hline
\end{tabular}
\end{center}
\end{table}

\section{Conclusions \& future work}
In this investigation Amazon's prototype SSWS system has been comprehensively tested against hybrid synthesis and conventional statistical parametric speech synthesis systems to demonstrate its performance. This perceptual testing has been repeated to show that its findings are reliable. To better understand the performance of SSWS more in-depth analysis has been conducted relating to the speech domain. In addition the nature of the speech errors between the different speech synthesis technologies has been analysed. From this analysis, we can see that SSWS reduces the number of speech errors in a number of different categories. It however suffers from audio glitches in the synthesised speech waveforms. 

Following this investigation the main priority is to improve the stability of SSWS to reduce the number of audio glitches produced. This investigation has also highlighted the importance of the system used to generate the prosody contour that conditions the SSWS system. Future work is needed to improve prosody modelling in order to enable SSWS to achieve greater naturalness across multiple domains.

\bibliographystyle{IEEEbib}
\bibliography{references}

\end{document}